\documentclass[aps,prl,twocolumn,superscriptaddress,floatfix]{revtex4}

\usepackage{graphics}
\usepackage{color}

\begin{document}

\title{Shape Coexistence at Zero Spin in  $^{64}$Ni Driven by the Monopole Tensor Interaction}



\author{N.~M\u{a}rginean\footnote{Corresponding author: Nicu@tandem.nipne.ro} }
\affiliation{Horia Hulubei National Institute of Physics and Nuclear Engineering - IFIN HH, Bucharest, 077125, Romania}
\author{D.~Little}
\affiliation{Dept. of Physics and Astronomy, University of North Carolina at Chapel Hill, Chapel Hill, North Carolina 27599-3255, USA }
\affiliation{Triangle Universities Nuclear Laboratory, Duke University, Durham, North Carolina 27708-2308, USA}
\author{Y.~Tsunoda}
\affiliation{Center for Nuclear Study, The University of Tokyo, 7-3-1 Hongo, Bunkyo, Tokyo 113-0033, Japan}
\author{S.~Leoni\footnote{Corresponding author: Silvia.Leoni@mi.infn.it} }
\affiliation{Dipartimento di Fisica, Universit$\grave{a}$ degli Studi di Milano, I-20133 Milano, Italy}
\affiliation{INFN sezione di Milano via Celoria 16, 20133, Milano, Italy }
\author{R. V. F.~Janssens\footnote{Corresponding author: Rvfj@email.unc.edu}}
\affiliation{Dept. of Physics and Astronomy, University of North Carolina at Chapel Hill, Chapel Hill, North Carolina 27599-3255, USA }
\affiliation{Triangle Universities Nuclear Laboratory, Duke University, Durham, North Carolina 27708-2308, USA}
\author{B.~Fornal\footnote{Corresponding author: Bogdan.Fornal@ifj.edu.pl}}
\affiliation{Institute of Nuclear Physics, PAN, 31-342 Krak\'ow, Poland }
\author{T.~Otsuka\footnote{Corresponding author: Otsuka@phys.s.u-tokyo.ac.jp} }
\affiliation{Department of Physics, The University of Tokyo, 7-3-1 Hongo, Bunkyo, Tokyo 113-0033, Japan }
\affiliation{RIKEN Nishina Center, 2-1 Hirosawa, Wako, Saitama 351-0198, Japan}
\affiliation{KU Leuven, Instituut voor Kern- en Stralingsfysica, 3000 Leuven, Belgium }
\author{C.~Michelagnoli}
\affiliation{Institut Laue-Langevin (ILL), 71 Avenue des Martyrs, 38042 Grenoble, France}
\author{L.~Stan}
\affiliation{Horia Hulubei National Institute of Physics and Nuclear Engineering - IFIN HH, Bucharest, 077125, Romania}
\author{F.~C.~L.~Crespi}
\affiliation{Dipartimento di Fisica, Universit$\grave{a}$ degli Studi di Milano, I-20133 Milano, Italy}
\affiliation{INFN sezione di Milano via Celoria 16, 20133, Milano, Italy }
\author{C.~Costache}
\affiliation{Horia Hulubei National Institute of Physics and Nuclear Engineering - IFIN HH, Bucharest, 077125, Romania}
\author{R.~Lica}
\affiliation{Horia Hulubei National Institute of Physics and Nuclear Engineering - IFIN HH, Bucharest, 077125, Romania}
\author{M.~Sferrazza}
\affiliation{D\'epartement de Physique, Universit\'e libre de Bruxelles, B-1050 Bruxelles, Belgium}
\author{A.~Turturica}
\affiliation{Horia Hulubei National Institute of Physics and Nuclear Engineering - IFIN HH, Bucharest, 077125, Romania}
\author{A.~D.~Ayangeakaa}
\affiliation{Department of Physics, United States Naval Academy, Annapolis, Maryland 21402, USA}
\author{K.~Auranen\footnote{Present address: Department of Physics, University of Jyvaskyla, P.O. Box 35, 40014, Jyvaskyla, Finland}}
\affiliation{Physics Division, Argonne National Laboratory, Argonne, Illinois 60439, USA}
\author{M.~Barani}
\affiliation{Dipartimento di Fisica, Universit$\grave{a}$ degli Studi di Milano, I-20133 Milano, Italy}
\affiliation{INFN sezione di Milano via Celoria 16, 20133, Milano, Italy }
\affiliation{Institut Laue-Langevin (ILL), 71 Avenue des Martyrs, 38042 Grenoble, France}
\author{P.~C.~Bender}
\affiliation{Department of Physics and Applied Physics, University of Massachusetts Lowell, Lowell, Massachusetts 01854, USA}
\author{S.~Bottoni}
\affiliation{Dipartimento di Fisica, Universit$\grave{a}$ degli Studi di Milano, I-20133 Milano, Italy}
\affiliation{INFN sezione di Milano via Celoria 16, 20133, Milano, Italy }
\author{M.~Boromiza}
\affiliation{Horia Hulubei National Institute of Physics and Nuclear Engineering - IFIN HH, Bucharest, 077125, Romania}
\author{A.~Bracco}
\affiliation{Dipartimento di Fisica, Universit$\grave{a}$ degli Studi di Milano, I-20133 Milano, Italy}
\affiliation{INFN sezione di Milano via Celoria 16, 20133, Milano, Italy }
\author{S.~C\u alinescu}
\affiliation{Horia Hulubei National Institute of Physics and Nuclear Engineering - IFIN HH, Bucharest, 077125, Romania}
\author{C.~M.~Campbell}
\affiliation{Nuclear Science Division, Lawrence Berkeley National Laboratory, Berkeley, California 94720, USA}
\author{M.~P.~Carpenter}
\affiliation{Physics Division, Argonne National Laboratory, Argonne, Illinois 60439, USA}
\author{P.~Chowdhury}
\affiliation{Department of Physics and Applied Physics, University of Massachusetts Lowell, Lowell, Massachusetts 01854, USA}
\author{M.~Ciema\l a}
\affiliation{Institute of Nuclear Physics, PAN, 31-342 Krak\'ow, Poland }
\author{N.~Cieplicka-Ory\`nczak}
\affiliation{Institute of Nuclear Physics, PAN, 31-342 Krak\'ow, Poland }
\author{D.~Cline}
\affiliation{Department of Physics and Astronomy, University of Rochester, Rochester, New York 14627, USA}
\author{C.~Clisu}
\affiliation{Horia Hulubei National Institute of Physics and Nuclear Engineering - IFIN HH, Bucharest, 077125, Romania}
\author{H.~L.~Crawford}
\affiliation{Nuclear Science Division, Lawrence Berkeley National Laboratory, Berkeley, California 94720, USA}
\author{I.~E.~Dinescu}
\affiliation{Horia Hulubei National Institute of Physics and Nuclear Engineering - IFIN HH, Bucharest, 077125, Romania}
\author{D.~Filipescu}
\affiliation{Horia Hulubei National Institute of Physics and Nuclear Engineering - IFIN HH, Bucharest, 077125, Romania}
\author{N.~Florea}
\affiliation{Horia Hulubei National Institute of Physics and Nuclear Engineering - IFIN HH, Bucharest, 077125, Romania}
\author{A.~M.~Forney}
\affiliation{Department of Chemistry and Biochemistry, University of Maryland, College Park, Maryland 20742, USA}
\author{S.~Fracassetti\footnote{Present address: KU Leuven, Instituut voor Kern- en Stralingsfysica, 3000 Leuven, Belgium}}
\affiliation{Dipartimento di Fisica, Universit$\grave{a}$ degli Studi di Milano, I-20133 Milano, Italy}
\affiliation{INFN sezione di Milano via Celoria 16, 20133, Milano, Italy }
\author{A.~Gade}
\affiliation{Department of Physics and Astronomy, Michigan State University, East Lansing, Michigan 48824, USA}
\affiliation{National Superconducting Cyclotron Laboratory, Michigan State University, East Lansing, Michigan 48824, USA}
\author{I.~Gheorghe}
\affiliation{Horia Hulubei National Institute of Physics and Nuclear Engineering - IFIN HH, Bucharest, 077125, Romania}
\author{A.~B.~Hayes}
\affiliation{National Nuclear Data Center, Brookhaven National Laboratory, Upton, New York 11973-5000, USA}
\author{I.~Harca}
\affiliation{Horia Hulubei National Institute of Physics and Nuclear Engineering - IFIN HH, Bucharest, 077125, Romania}
\author{J.~Henderson}
\affiliation{Lawrence Livermore National Laboratory, Livermore, California 94550, USA}
\author{A.~Ionescu}
\affiliation{Horia Hulubei National Institute of Physics and Nuclear Engineering - IFIN HH, Bucharest, 077125, Romania}
\author{\L.~W.~Iskra}
\affiliation{INFN sezione di Milano via Celoria 16, 20133, Milano, Italy }
\author{M.~Jentschel}
\affiliation{Institut Laue-Langevin (ILL), 71 Avenue des Martyrs, 38042 Grenoble, France}
\author{F.~Kandzia}
\affiliation{Institut Laue-Langevin (ILL), 71 Avenue des Martyrs, 38042 Grenoble, France}
\author{Y.~H.~Kim}
\affiliation{Institut Laue-Langevin (ILL), 71 Avenue des Martyrs, 38042 Grenoble, France}
\author{F.~G.~Kondev}
\affiliation{Physics Division, Argonne National Laboratory, Argonne, Illinois 60439, USA}
\author{G.~Korschinek}
\affiliation{Technische Universitat Munchen, Munchen, Germany}
\author{U.~K\"oster}
\affiliation{Institut Laue-Langevin (ILL), 71 Avenue des Martyrs, 38042 Grenoble, France}
\author{Krishichayan}
\affiliation{Triangle Universities Nuclear Laboratory, Duke University, Durham, North Carolina 27708-2308, USA}
\author{M.~Krzysiek}
\affiliation{Institute of Nuclear Physics, PAN, 31-342 Krak\'ow, Poland }
\author{T.~Lauritsen}
\affiliation{Physics Division, Argonne National Laboratory, Argonne, Illinois 60439, USA}
\author{J.~Li\footnote{Present address: Department of Physics and Astronomy and National Superconducting Cyclotron Laboratory, Michigan State University, East Lansing, Michigan 48824, USA}}
\affiliation{Physics Division, Argonne National Laboratory, Argonne, Illinois 60439, USA}
\author{R.~M\u{a}rginean}
\affiliation{Horia Hulubei National Institute of Physics and Nuclear Engineering - IFIN HH, Bucharest, 077125, Romania}
\author{E.~A.~Maugeri}
\affiliation{Paul Scherrer Institut, Villigen, Switzerland}
\author{C.~Mihai}
\affiliation{Horia Hulubei National Institute of Physics and Nuclear Engineering - IFIN HH, Bucharest, 077125, Romania}
\author{R.~E.~Mihai}
\affiliation{Horia Hulubei National Institute of Physics and Nuclear Engineering - IFIN HH, Bucharest, 077125, Romania}
\author{A.~Mitu}
\affiliation{Horia Hulubei National Institute of Physics and Nuclear Engineering - IFIN HH, Bucharest, 077125, Romania}
\author{P.~Mutti}
\affiliation{Institut Laue-Langevin (ILL), 71 Avenue des Martyrs, 38042 Grenoble, France}
\author{A.~Negret}
\affiliation{Horia Hulubei National Institute of Physics and Nuclear Engineering - IFIN HH, Bucharest, 077125, Romania}
\author{C.~R.~Ni\c t\u a}
\affiliation{Horia Hulubei National Institute of Physics and Nuclear Engineering - IFIN HH, Bucharest, 077125, Romania}
\author{A.~Ol\u acel}
\affiliation{Horia Hulubei National Institute of Physics and Nuclear Engineering - IFIN HH, Bucharest, 077125, Romania}
\author{A.~Oprea}
\affiliation{Horia Hulubei National Institute of Physics and Nuclear Engineering - IFIN HH, Bucharest, 077125, Romania}
\author{S.~Pascu}
\affiliation{Horia Hulubei National Institute of Physics and Nuclear Engineering - IFIN HH, Bucharest, 077125, Romania}
\author{C.~Petrone}
\affiliation{Horia Hulubei National Institute of Physics and Nuclear Engineering - IFIN HH, Bucharest, 077125, Romania}
\author{C.~Porzio}
\affiliation{Dipartimento di Fisica, Universit$\grave{a}$ degli Studi di Milano, I-20133 Milano, Italy}
\affiliation{INFN sezione di Milano via Celoria 16, 20133, Milano, Italy }
\author{D.~Rhodes}
\affiliation{Department of Physics and Astronomy, Michigan State University, East Lansing, Michigan 48824, USA}
\affiliation{National Superconducting Cyclotron Laboratory, Michigan State University, East Lansing, Michigan 48824, USA}
\author{D.~Seweryniak}
\affiliation{Physics Division, Argonne National Laboratory, Argonne, Illinois 60439, USA}
\author{D.~Schumann}
\affiliation{Paul Scherrer Institut, Villigen, Switzerland}
\author{C.~Sotty}
\affiliation{Horia Hulubei National Institute of Physics and Nuclear Engineering - IFIN HH, Bucharest, 077125, Romania}
\author{S.~M.~Stolze}
\affiliation{Physics Division, Argonne National Laboratory, Argonne, Illinois 60439, USA}
\author{R.~ \c Suv\u{a}il\u{a}}
\affiliation{Horia Hulubei National Institute of Physics and Nuclear Engineering - IFIN HH, Bucharest, 077125, Romania}
\author{S.~Toma}
\affiliation{Horia Hulubei National Institute of Physics and Nuclear Engineering - IFIN HH, Bucharest, 077125, Romania}
\author{S.~Ujeniuc}
\affiliation{Horia Hulubei National Institute of Physics and Nuclear Engineering - IFIN HH, Bucharest, 077125, Romania}
\author{W.~B.~Walters}
\affiliation{Department of Chemistry and Biochemistry, University of Maryland, College Park, Maryland 20742, USA}
\author{C.~Y.~Wu}
\affiliation{Lawrence Livermore National Laboratory, Livermore, California 94550, USA}
\author{J.~Wu}
\affiliation{Physics Division, Argonne National Laboratory, Argonne, Illinois 60439, USA}
\author{S.~Zhu}
\affiliation{National Nuclear Data Center, Brookhaven National Laboratory, Upton, New York 11973-5000, USA}
\author{S.~Ziliani}
\affiliation{Dipartimento di Fisica, Universit$\grave{a}$ degli Studi di Milano, I-20133 Milano, Italy}
\affiliation{INFN sezione di Milano via Celoria 16, 20133, Milano, Italy }

\date{\today}

\begin{abstract}

The low-spin structure of the semi-magic $^{64}$Ni nucleus has been considerably expanded: combining four experiments, several 0$^+$ and 2$^+$ excited states were identified below 4.5 MeV, and their properties established. The Monte Carlo shell model accounts for the results and unveils an unexpectedly-complex landscape of coexisting shapes: a prolate 0$^+$ excitation is located at a surprisingly high energy (3463 keV), with a collective 2$^+$ state 286 keV above it, the first such observation in Ni isotopes. The evolution in excitation energy of the prolate minimum across the neutron $N=40$ sub-shell gap highlights the impact of the monopole interaction and its variation in strength with $N$.
\end{abstract}


\maketitle

In mesoscopic systems with many degrees of freedom (e.g., molecules, atomic nuclei, etc.), deformation is a common phenomenon resulting from symmetry breaking associated with quantum-mechanical states (practically) degenerate in energy. The concept was originally introduced by Jahn and Teller who demonstrated that, in non-linear molecules, coupling between degenerate electronic states and collective vibrations can destroy the system's original symmetry \cite{JT37}.  In atomic nuclei, the appearance of ellipsoidal deformation is a realization of this effect with specific superpositions of spherical single-particle states (e.g., Nilsson model \cite{BM-2}) induced by deformed mean potentials (mean-field approaches) \cite{Rei84,Naz94}, or by quadrupole correlations (shell-model descriptions) \cite{Uts12,Ots19}, highlighting the interplay between single-particle states and collective modes.

Among the features associated with deformation figures shape coexistence: a phenomenon ubiquitous throughout the nuclear chart \cite{Woo92,Hey11} where different shapes are present at comparable excitation energies. A clear-cut signature for its occurrence in even-even systems is the presence of  low-lying 0$^+$ excitations residing in local minima of the nuclear potential energy surface (PES) in deformation space.

Over the past two decades, studies of neutron-rich nuclei have highlighted 
the contribution of the monopole component of the tensor force to the evolution of the structure of exotic nuclei
\cite{Ots10,Ots19}, especially in the change in single-particle (or shell) structure with neutron excess, with some magic numbers vanishing and other, new ones appearing  \cite{Ots20}. Besides such single-particle properties, its role in driving the nuclear shape was subsequently identified \cite{Tsu14,Ots16,Tog16,Mar18,Sel19,Ots19}, specifically in connection with shape coexistence.

Neutron-rich even $_{28}$Ni isotopes are a noteworthy example of shape coexistence:
$^{68}$Ni exhibits a spherical ground state, while the 1605-keV, 0$^+_2$ and 2511-keV, 0$^+_3$  levels  are understood as oblate and prolate excitations \cite{Tsu14,Cri16,Suc14,Rec13,Fla15,Mue00,Chi12}. In $^{70}$Ni, a prolate 
0$^+_2$ state is found at 1567 keV above the spherical ground state \cite{Pro15}. Finally, four 0$^+$ levels are known below a 4 MeV excitation energy in $^{66}$Ni,  where the ground state and the 2664-keV 0$^+_3$ level are interpreted as spherical, while the 0$^+_2$, 2445-keV and the 0$^+_4$, 2945-keV states are of oblate and prolate character \cite{Leo17}.

The present paper focuses on $^{64}$Ni, the heaviest, stable nucleus in the isotopic chain, and reveals a complex landscape in deformation that was not anticipated by mean-field calculations  \cite{Hil07,Ber91,Dec80,Mol04,Mol09}, which predicted a single, spherical minimum, the development of a secondary prolate one occurring only in heavier isotopes. In contrast, recent Monte Carlo Shell-model (MCSM) calculations \cite{Tsu14} indicate coexistence of spherical and deformed oblate and prolate 0$^+$ states already in $^{62,64}$Ni. This coexistence originates from the action of the monopole tensor force which shifts effective single-particle energies, already at the valley of stability, weakening resistance against deformation \cite{Tsu14, Ots16, Ots19}. This Letter reports extensive tests of these MCSM predictions. Besides the customary data on level energies, spins and parities, comparisons also extend to state lifetimes, transition probabilities, branching and multipole mixing ratios. Evidence is given for three coexisting shapes, with the prolate 0$^+$ state at $\sim$3.5 MeV, an excitation energy reproduced only by MCSM calculations incorporating the monopole tensor interaction. To achieve the required experimental sensitivity, four experiments; i.e., transfer reactions, neutron capture, Coulomb excitation and nuclear resonance fluorescence  had to be carried out at the IFIN-HH Tandem Laboratory (Bucharest, Romania), the Institut Laue-Langevin (ILL, Grenoble, France), the Argonne National Laboratory (ANL, Argonne, USA) and the Triangle Universities Nuclear Laboratory (TUNL, Duke Univ., USA), respectively. Results from the first three techniques are reported below (for the last one, see Ref. \cite{Fri21}). 

Prior to this work, two excited 0$^+$ states had been identified in $^{64}$Ni, at 2867 and 3026 keV, following $\beta$-decay and (t,p)-reaction studies  \cite{Paw12,Dar71}. These levels were subsequently confirmed in deep-inelastic reaction measurements \cite{Bro12}, and their $\gamma$ decay to the 1346-keV 2$^+_1$ state was observed. Candidates for other, higher-lying 0$^+$ levels have also been reported \cite{Har92}. 

At IFIN-HH, $^{64}$Ni was populated by $^{62}$Ni($^{18}$O,$^{16}$O) two-neutron (2n) transfer on a 5 mg/cm$^2$-thick target, with a 39-MeV beam energy; i.e., just below the Coulomb barrier in order to reduce competition from fusion-evaporation. Transitions of interest were measured with ROSPHERE, an array of 25 Compton-suppressed HPGe detectors with $\sim$2$\%$ total efficiency at 1.3 MeV \cite{Buc16}. The same reaction, but with a thinner, 0.92 mg/cm$^2$ target and a 5 mg/cm$^2$-Ta stopper, placed at six distances from the target (10, 17, 25, 45, 100 and 150 $\mu$m), was employed for lifetime measurements via the recoil-distance technique. The sub-barrier one-proton (1p) transfer reaction $^{65}$Cu($^{11}$B,$^{12}$C)$^{64}$Ni at 26 MeV on a 7.22 mg/cm$^2$-thick target was  performed as well \cite{Stan20}.

\begin{figure}[ht]
\resizebox{0.48\textwidth}{!}{\includegraphics{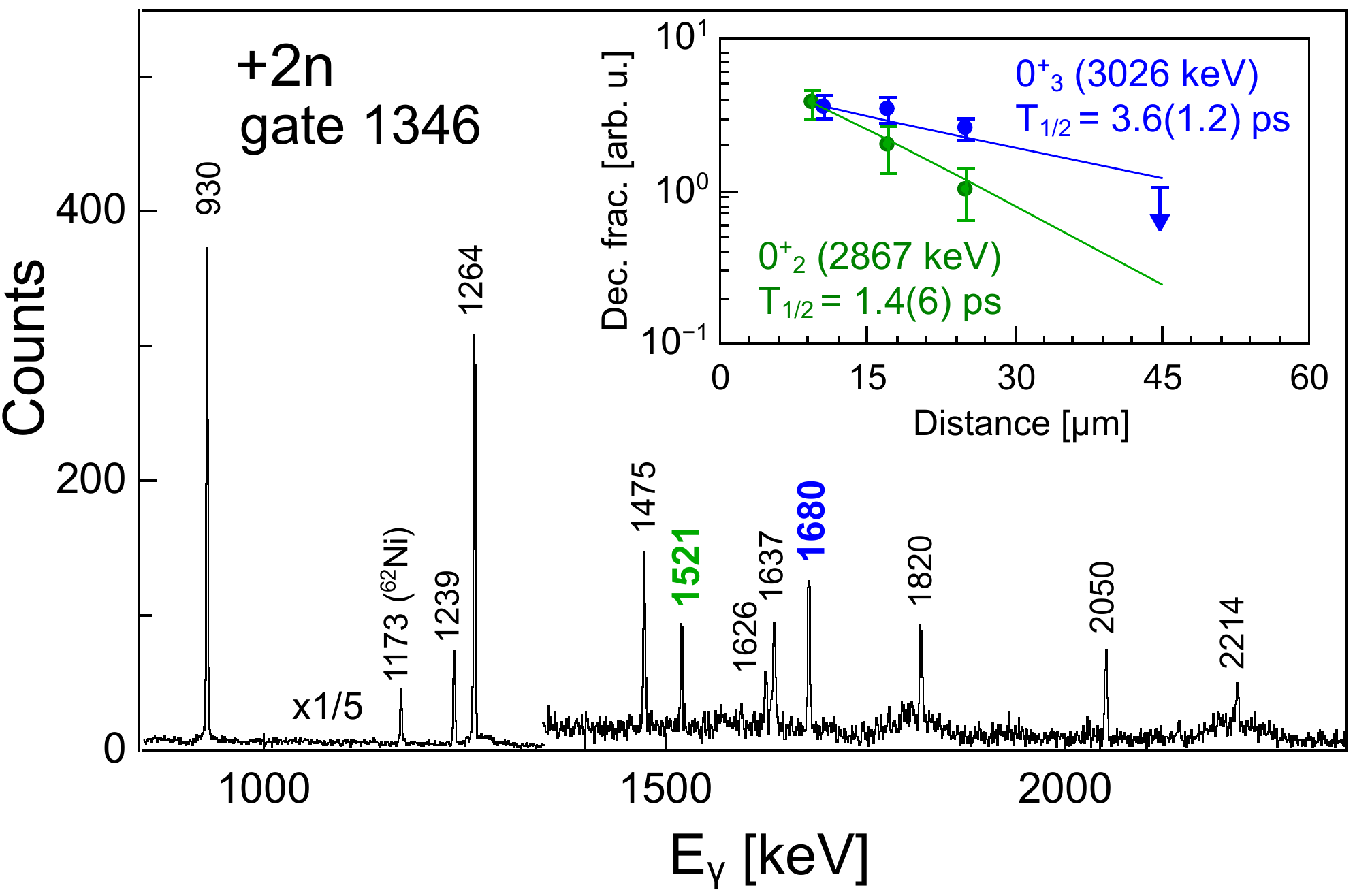}}
\caption{2n-transfer reaction: partial spectrum  in coincidence with the $^{64}$Ni $2^+_1 \rightarrow$0$^+_1$ transition (thick-target data). Inset: decay curves from the plunger experiment for the 1521- and 1680-keV transitions linking the 0$^+_2$ and 0$^+_3$ levels to the 2$^+_1$ state. } 
\label{plunger}
\end{figure}

The coincidence spectrum, from the thick target 2n-transfer reaction, gated on the 1346-keV 2$^+_1 \rightarrow$0$^+_1$ $^{64}$Ni transition, is presented in Fig. \ref{plunger}. All visible lines correspond to $^{64}$Ni transitions which deexcite states below 4.6 MeV, with spin up to 7$^-$ \cite{Bro12}. The 1521- and 1680-keV $\gamma$ rays, depopulating the known 0$^+_2$ and 0$^+_3$ states, are clearly visible. Their respective half-lives, as measured in the plunger experiment, are $T_{1/2}$=1.4(6) and 3.6(1.2) ps (inset of Fig. \ref{plunger})  \cite{Stan20}.  A search for transitions from higher-lying candidate 0$^+$ states proved inconclusive in the IFIN-HH data sets.

A twenty-day experiment was then conducted at ILL \cite{Mar20}, where $^{64}$Ni was populated via thermal-neutron capture on a 2~GBq $^{63}$Ni 
sample (extracted from a larger CERN-nTOF sample \cite{Led14}), where  12.1 mg of NiO grains were glued between two 6~$\mu$m-thick Al foils and contained $\approx 8$\% $^{63}$Ni (T$_{1/2}$ = 101.2(15) y), 69\% $^{62}$Ni, $<$3\% $^{63}$Cu and other Ni isotopes, and 20\% O.  The $\gamma$ rays were detected with FIPPS  \cite{Mic18}, a $\sim$3.7$\%$ efficient array of 16 clover detectors (8 on loan from IFIN-HH) arranged in a 4$\pi$ geometry, particularly suitable for angular correlation studies in view of the large number (71) of relative angles between the germanium crystals.

\begin{figure}[ht]
\resizebox{0.47\textwidth}{!}{\includegraphics{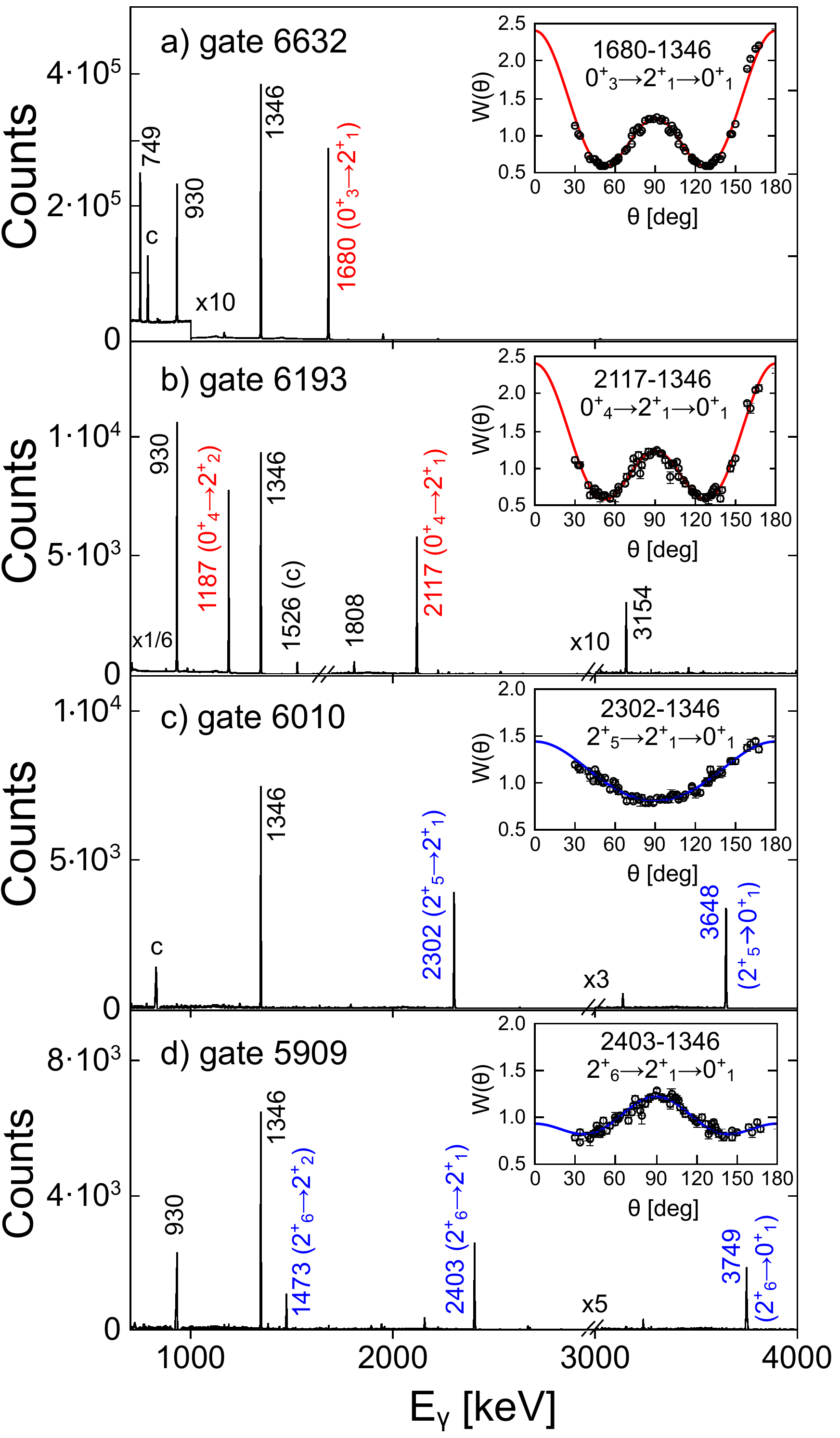}}
\caption{Neutron-capture data: $\gamma$-ray spectra in coincidence with primary transitions (energy  indicated in each panel).  Relevant transitions are also labeled. Insets: measured angular correlations for the  cascades indicated.}
\label{ang-cor}
\end{figure}

The 1/2$^-$ spin-parity of the $^{63}$Ni ground state results in a 0$^-$ or 1$^-$ $^{64}$Ni capture state at 9657.47 keV, and strong population of 0$^+$, 1$^+$ and 2$^+$ levels through $E1$ primary $\gamma$ rays is expected. Direct feeding of the known 0$^+_2$ and 0$^+_3$ states, at 2866.9 and 3025.5 keV, is clearly visible in coincidence spectra gated on respective 6791- and 6632-keV primary transitions. Figure \ref{ang-cor}(a) displays the 6632-keV gated spectrum where both the 0$^+_3\rightarrow$ 2$^+_1$, 1680-keV transition and a new, weaker  (3.6(2)$\%$ of the former) 749-keV, 0$^+_3\rightarrow$ 2$^+_2$ decay branch are observed. The angular correlation for the 0$^+_3\rightarrow$2$^+_1 \rightarrow$ 0$^+_1$ cascade (inset, Fig. \ref{ang-cor}(a)) agrees with the 0$^+$ spin-parity assignment to the 3026-keV level. A search for additional 0$^+$ excited states was undertaken by considering every primary transition in coincidence
with the 1346-keV ground-state transition, and also feeding levels in the 3-6 MeV excitation energy range. Five such high-energy transitions, at 6194, 5801, 5389, 4954 and 3889 keV, populating states at 3463.1, 3856.0, 4268.1, 4703.9 and 5768.6 keV were found to exhibit decay patterns only consistent with 0$^+$ spin-parity assignments \footnote{For the 3856.0-keV level, the 0$^+$ assignment was established via the 702-3154 keV correlation cascade, involving the intermediate 1$^+$ state at 3154 keV, first identified in this work  \cite{Mar20}).}. Further, the angular correlation analysis yielded firm 0$^+$ assignments for the states at 3463.1, 4268.1, 4703.9 and 5768.6 keV by considering in each case pairs of $\gamma$ rays composed, on the one hand, of the decay branch to the 2$^+_1$ state and, on the other, of the 2$^+_1 \rightarrow$0$^+_1$, 1346-keV transition. The relevant analysis for the 0$^+_4$, 3463.1-keV level is illustrated in Fig. \ref{ang-cor}(b).

\begin{figure*}[ht]
\resizebox{1.0\textwidth}{!}{\includegraphics{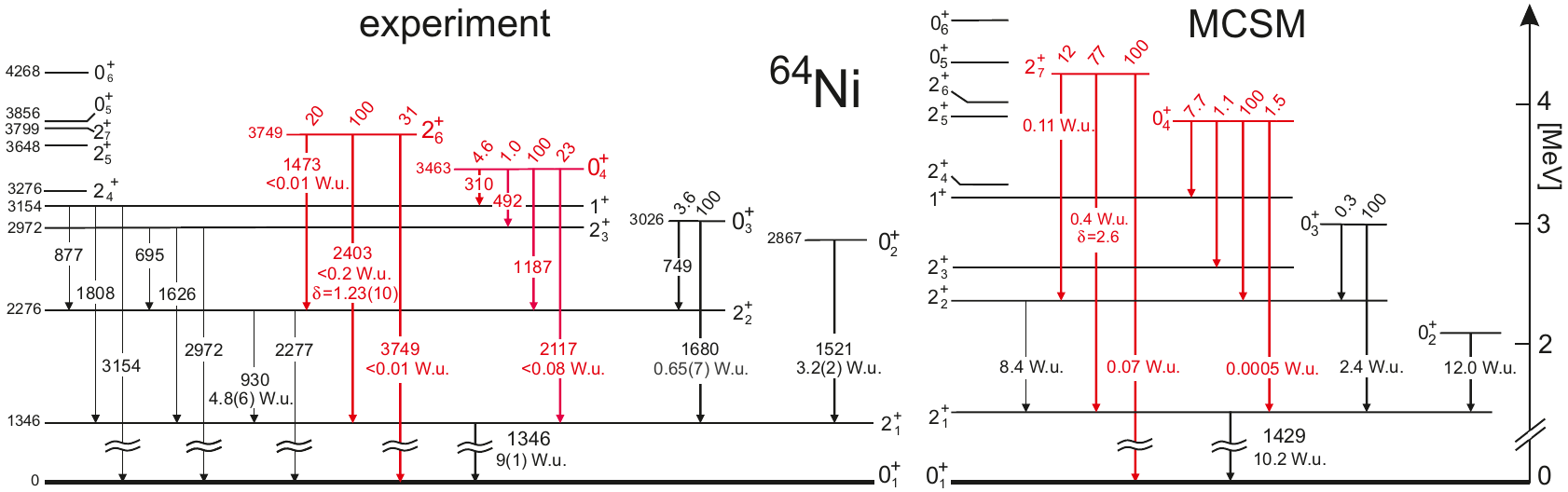}}
\caption{Data (left) and MCSM calculations (right) for $^{64}$Ni level properties derived in this work. Transition strengths are from CE, branchings and mixing ratios from neutron capture.  In red, decays from prolate structures.}
\label{level-scheme}
\end{figure*}

The partial level scheme is given in Fig. \ref{level-scheme}. The 0$^+_4$ state at 3463 keV is of particular interest; it was observed earlier with tentative (2$^{+/-}$, 3$^-$)  \cite{Bro12} and (0$^+$, 1, 2, 3$^-$)  spin-parity assignments \cite{Har92}, but is firmly assigned here. It should be emphasized that this level is not populated in $^{64}$Co $\beta$-decay  \cite{Paw12}, in contrast to all other 0$^+$ states, up to 0$^+_6$, which are fed in this process. This observation already points to a marked difference in structure for this excitation, and is reminiscent of that occurring in $^{66}$Ni  \cite{Leo17},  where the prolate-deformed 0$^+_4$ state at 2974 keV was also the only 0$^+$ excitation not fed in the $\beta$-decay of the spherical $^{66}$Co ground state  \cite{Str18}.
Further inspection of the ILL data revealed three 2$^+$ states (firmly established in this work) at  3647.9, 3749.1 and 3798.7 keV, which complement four such excitations, at 1345.8, 2276.6, 2972.1 and 3276.0 keV, reported in Refs.  \cite{NUDAT} (see Fig. \ref{level-scheme}). The angular correlations for the transitions deexciting 2$^+_4$, 2$^+_5$ and 2$^+_7$ levels toward the 2$^+_1$ state all indicate a dominant $M1$ character, with only a small $E2$ admixture. This is illustrated through the representative data for the 2$^+_5 \rightarrow$ 2$^+_1 \rightarrow$ 0$^+_1$ cascade of Fig. \ref{ang-cor}(c). A notable exception to this trend is the 2$^+_6 \rightarrow$ 2$^+_1 \rightarrow$0$^+_1$ sequence where the 2403-keV $\gamma$ ray exhibits a pronounced $E2$ character, as strikingly illustrated by comparing Figs. \ref{ang-cor}(c) and (d). The corresponding mixing ratio was determined to be $\delta$ = +1.23(10). Furthermore, from the line shape observed for this 2403-keV $\gamma$ ray in spectra following 1p transfer, a lower limit of 0.5 ps was obtained for the 2$^+_6$ state half-life, which results in upper limits to the respective $B$($E2$) strengths of 0.02, 0.4, and 0.02 W.u. for the 2$^+_6 \rightarrow$ 0$^+_1$, 2$^+_6 \rightarrow$2$^+_1$ and 2$^+_6 \rightarrow$  2$^+_2$ transitions. These all indicate that deexcitation from the 2$^+_6$ state is significantly hindered.

To gain further insight into the nature of the 0$^+$ and 2$^+$ excitations in $^{64}$Ni, a multi-step Coulomb excitation (CE) measurement was carried out at the ATLAS facility at ANL. A 0.5 mg/cm$^2$ isotopically-enriched $^{208}$Pb target was bombarded by a $^{64}$Ni beam of 272 MeV, a value 14$\%$ below the Coulomb barrier. Transitions were detected with the GRETINA tracking array  \cite{Pas13} in coincidence with the two reaction partners measured by the CHICO2 compact heavy ion counter  \cite{Wu16}, as described in Ref.  \cite{Aya19}.
The resulting yields were analyzed with the semi-classical Coulomb excitation code, GOSIA  \cite{Czo12}, which provided electromagnetic matrix elements for transitions from 13 states \cite{Lit20}. For the present paper, the following reduced transition probabilities are relevant: $B$($E2$; 2$^+_1 \rightarrow$0$^+_1$) = 140(20) $e^2$fm$^4$, $B$($E2$;2$^+_2 \rightarrow$  2$^+_1$) = 73(8) $e^2$fm$^4$,  $B$($E2$; 0$^+_2 \rightarrow$ 2$^+_1$) = 48(3) $e^2$fm$^4$, $B$($E2$; 0$^+_3 \rightarrow$ 2$^+_1$) = 10(1) $e^2$fm$^4$, $B$($E2$; 0$^+_4 \rightarrow$ 2$^+_1$ ) $<$ 1.3 $e^2$fm$^4$ and $B$($E2$; 2$^+_6  \rightarrow$ 2$^+_2$) $<$ 3.2 $e^2$fm$^4$. In terms of single-particle estimates, these values translate into strengths of 9(1), 4.8(5), 3.2(2), 0.65(7), $<$0.08 and $<$0.2 W.u., respectively. Good agreement is noted with the lifetime data for the 0$^+_2$ and 0$^+_3$ decays to the 2$^+_1$
state with $B$($E2$) values of 3.3(14) and 0.8(3) W.u., respectively (Fig. \ref{plunger}). The limits for the 2$^+_6$ and 0$^+_4$ states are also consistent, but somewhat tighter in the CE case.

The left part of Fig. \ref{level-scheme} provides a $^{64}$Ni level scheme encompassing states of positive parity with spins 0, 1, 2, up to 4.26 MeV excitation energy. The information gathered on mixing, branching ratios and transition strengths has been added to enable extensive comparisons with calculations. Such additional tests of theory are generally difficult, if not impossible, especially for neutron-rich nuclei requiring experiments with radioactive beams.  


The right side of Fig. \ref{level-scheme} presents the level scheme from MCSM calculations, performed with significantly extended MCSM basis vectors as compared to earlier studies of $^{66-78}$Ni  \cite{Leo17,Tsu14}. The model space includes protons and neutrons in the full $fp$ shell with, in addition, the $g_{9/2}$ and $d_{5/2}$ orbitals, and the Hamiltonian is based on the A3DA-m effective interaction \cite{Tsu14}. The transition probabilities were obtained with standard effective charges ($e_p$ = 1.5 $e$, $e_n$ = 0.5 $e$), a spin quenching factor of  0.7 and an isovector orbital $g$-factor of  0.1 \cite{Hon09}. State energies are reproduced satisfactorily $-$  the rms deviation is $\sim$300 keV, commensurate with expectations for shell-model calculations. For the first three 0$^+$ excitations, the computed decay patterns mirror the data:  the decay to the 2$^+_1$ level dominates the deexcitation from 0$^+_2$ and 0$^+_3$ states, and the branching ratios between the four transitions from the 0$^+_4$ state are qualitatively reproduced, 
with the 0$^+_4 \rightarrow$ 2$^+_2$ one being strongest. The relative $B$($E2$) strengths calculated for the 0$^+_{2,3,4}\rightarrow$ 2$^+_1$ decays (i.e., 12, 2.4 and 5 10$^{-4}$ W.u.) are consistent with the data, even though the absolute strengths are larger. Finally, the MCSM calculations also reproduce the lack of feeding of the 0$^+_4$ state in $\beta$ decay, when compared to that of the other 0$^+$ levels.
 
A sequence of relatively close-lying 2$^+$ levels is also predicted with deexcitation patterns and transition probabilities agreeing with observations, at least when the calculated 2$^+_7$ state is associated with the 2$^+_6$ experimental one  $ - $ the 2$^+_6$ and 2$^+_7$ levels are computed to lie only 235 keV apart; i.e., within the expected accuracy of the A3DA-m interaction.
Theory also reproduces (i) relative variations in $B$($E2$) values between the 2$^+$ levels (including the  retardation for the transitions out of the 2$^+_6$ state, which agrees with the observed small $B$($E2$) upper limits), and (ii) the strong $E2$ component in the $\Delta I$ = 1, 2$^+_6 \rightarrow$  2$^+_1$ transition, where the measured mixing ratio $\delta$($E2/M1$) = +1.23(10) (vs. $\delta_{\rm MCSM}$=2.6) contrasts those for similar transitions from the other 2$^+$ excitations (Fig. \ref{ang-cor} (c)). 

\begin{figure}[ht]
\resizebox{0.48\textwidth}{!}{\includegraphics{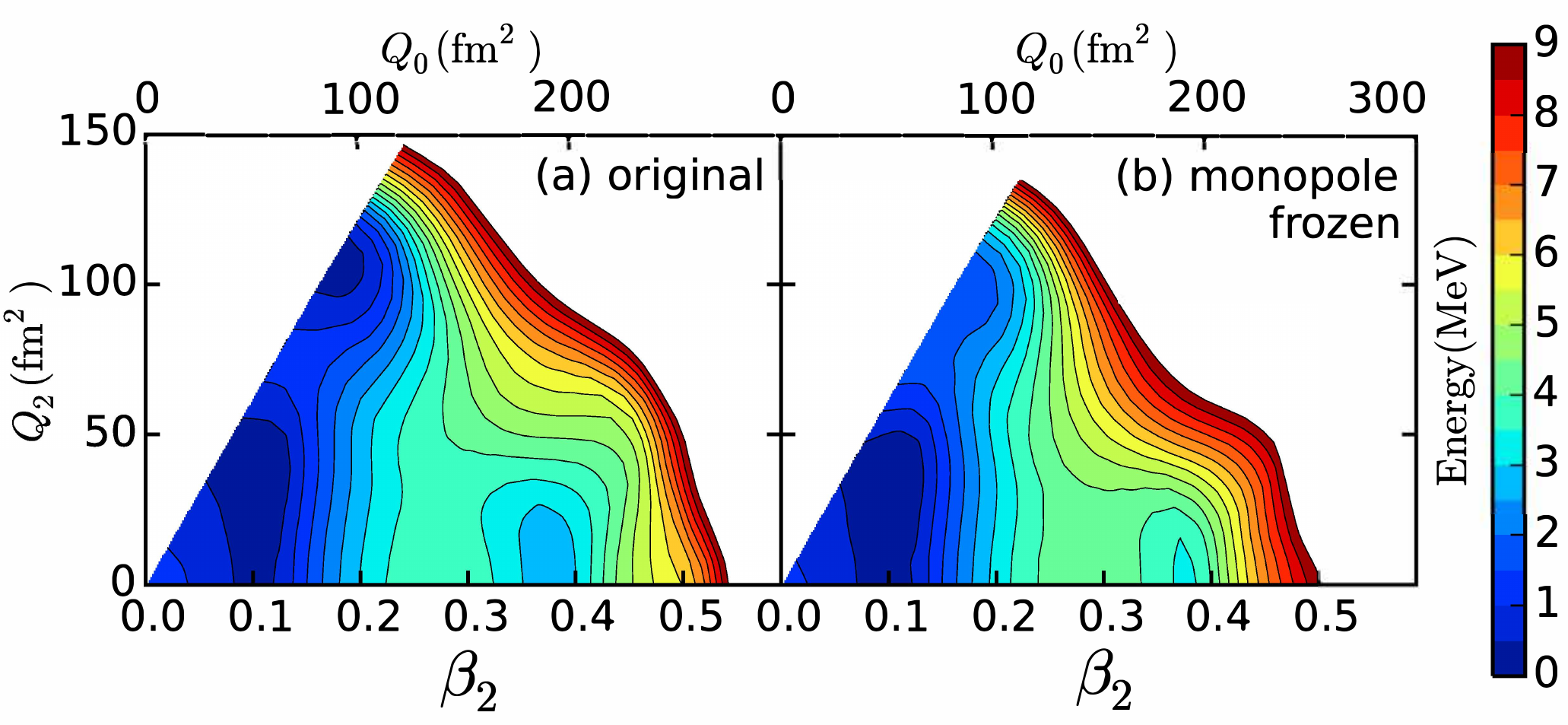}}
\caption{$^{64}$Ni potential energy surfaces with (a) full, original interaction 
used in MCSM calculations \cite{Tsu14}, and (b) monopole-frozen interaction 
(i.e., the monopole component is subtracted from 
the proton-neutron interaction, and single-particle energies are 
adjusted to original effective values of the spherical minimum \cite{Ots19}). } 
\label{Monopole-PES}
\end{figure}

According to the MCSM calculations,  the first four 0$^+$ states reside in spherical, oblate, spherical and prolate minima, respectively, in the PES obtained for the A3DA-m Hamiltonian by the constrained Hartree-Fock method \cite{Tsu14,Ots16}. Thus, the 0$^+_4 \rightarrow$2$^+_1$ decay is a prolate-to-spherical shape-changing transition, resulting in significant retardation, in line with the $B$($E2$) limit of $<$0.08 W.u. The same picture applies to the computed 2$^+_7$ level, which theory also locates in the prolate minimum. The observed decay pattern, the limits on the decay strengths and the dominant $E2$ character of the 2$^+_6 \rightarrow$ 2$^+_1$ transition argue in favor of this interpretation for the observed 2$^+_6$ state. Hence, based on the consistency between data and theory, this  2$^+_6$ level represents the first observation in the Ni isotopes of a 2$^+$ excited state in a well-isolated prolate potential minimum. The ``shape-isomer''-like properties of the 0$^+_4$ excitation in $^{64}$Ni mirror closely those found in $^{66}$Ni  \cite{Leo17} with, in addition, the observation of the first element of a rotational sequence. Using the Raman systematics, linking 2$^+$ energies to transition strengths \cite{Ram01}, this results in a computed $\beta_2$ deformation of $\sim$0.4, in agreement with the MCSM result 
(Fig. \ref{Monopole-PES}(a)). The low-energy 286-keV in-band 2$^+_6 \rightarrow$ 0$^+_4$ transition, even with a calculated $B$($E2$) strength of $\sim$40 W.u., cannot be observed: the flux proceeds through high-energy ($>$1 MeV)  $E2$ $\gamma$ rays due to the $E_{\gamma}^5$  factor.

With these new, extensive data in $^{64}$Ni, the evolution in energy of the prolate minimum with $N$ can now be traced in the Ni isotopes, revealing a sharp contrast with that exhibited by the 2$^+_1$ levels of spherical nature. The latter are all in the 1250 - 1450 keV range, with the exception of the 2033-keV value for $^{68}$Ni due to  the $N=40$ sub-shell closure.  In contrast, the prolate 0$^+$ excitation rises from 1567 keV in $^{70}$Ni, to 2511 and 2905 keV in $^{68}$Ni  and $^{66}$Ni, and 3463 keV in $^{64}$Ni. This behavior for $N<40$ differs markedly from the lowering of deformed intruder states when moving away from a (sub)-shell closure, observed in the Hg and Pb nuclei \cite{Woo92,Hey11}, for example. Low-lying prolate intruder states in the aforementioned neutron-rich Ni isotopes reflect the action of the monopole tensor force which is often referred to as Type II shell evolution \cite{Ots16,Tog16,Leo17},  and involves particle-hole excitations of neutrons to the $g_{9/2}$ unique-parity orbital from the $fp$ shell. Extra binding for such intruder states is provided largely by the monopole tensor part of the nucleon-nucleon force (the proton $f_{5/2}$ - $f_{7/2}$ spin-orbit splitting is reduced, favoring proton excitations across the $Z=28$ shell gap), which stabilizes isolated, deformed local minima in the PES (Fig. \ref{Monopole-PES}(a)). This additional binding is reduced for lower $N$ values as there are progressively fewer neutrons which can be excited to the $g_{9/2}$ orbital. The deformed minimum rises in excitation energy as a result. As demonstrated in Fig. \ref{Monopole-PES}(b), by deactivating  components of the monopole interaction  (i.e., monopole frozen \cite{Ots19}), a nearly vanishing prolate minimum would reside at even higher excitation, in line with mean field predictions \cite{Hil07,Ber91,Dec80,Mol04,Mol09}.

The present work has unveiled an unexpectedly complex landscape of nuclear deformation at zero spin in stable, semi-magic $^{64}$Ni. This includes the first identification, in Ni isotopes, of a  2$^+$ excitation in the prolate minimum.
The new results provide, for the first time, a complete picture of the mechanisms underlying the appearance of deformation and shape coexistence in the Ni isotopes. They highlight the impact of the monopole tensor interaction in driving deformation at zero spin, even in $^{64}$Ni, a nucleus within the valley of stability.



\bigskip

This work was partially supported by the European project for Nuclear Physics ENSAR2 (project number: 654002), by the Italian Istituto Nazionale di Fisica Nucleare, by the Polish National Science Centre under Contract No. 2014/14/M/ST2/00738 and 2013/08/M/ST2/00257 and by the Fonds de la Recherche Scientifique – FNRS Grant Number 4.45.10.08. It was supported by the Romanian Nucleu Project PN-19-06-01-02  and IFA-CERN-RO-06-ISOLDE. The MCSM calculations were performed on the K computer at RIKEN AICS and Oakforest-PACS operated by JCAHPC (hp160211, hp170230, hp180179, hp190160, hp200130). This work was supported in part by MEXT as ``Priority Issue on Post-K computer" (Elucidation of the Fundamental Laws and Evolution of the Universe)  and ``Program for Promoting Research on the Supercomputer Fugaku" (Simulation for basic science: from fundamental laws of particles to creation of nuclei)  and by JICFuS. 
This work was funded by the U.S. Department of Energy, Office of Science, Office of Nuclear Physics, under Contracts No. DE-AC02-06CH11357 (ANL), DE-AC52-07NA27344 (LLNL), DE-AC02-05CH11231 (LBNL), and under Grants No. DE-FG02-97ER41041 (UNC), No. DE-FG02-97ER41033 (TUNL), No. DE-FG02-08ER41556 and DE-SC0020451 (MSU), No. DE-FG02-94ER40848 (UML), and No. DE-FG02-94ER4084 (Maryland) and by the National Science Foundation under Grants No. PHY-1565546 (MSU) and No. PHY-1502092 (USNA). GRETINA was funded by the U.S. DOE, Office of Science, Office of Nuclear Physics under the ANL and LBNL contract numbers above. A. D. A. acknowledges support by Office of Naval Research (ONR) through the Naval Academy Research Council (NARC). This research used resources of ANLs ATLAS facility, a DOE Office of Science User Facility. The open access fee was covered by FILL2030, a European Union project within the European Commission’s Horizon 2020 Research and Innovation program under grant agreement No. 731096.


\begin{thebibliography}{20}
\bibitem{JT37} H.~A. Jahn and E.~Teller, Proc. Roy. Soc. A161, 220 (1937).
\bibitem{BM-2} A.~Bohr, B.~R.~Mottelson, Nuclear Structure, I and II, W.A. Benjamin, 1975.
\bibitem{Rei84} P.~-G.~Reinhard and E.~W.~Otten, Nucl. Phys. A420, 173 (1984).
\bibitem{Naz94} W.~Nazarewicz, Nuc. Phys. AA574, 27c (1994).
\bibitem{Uts12} Y.~Utsuno, T.~Otsuka, B.~A.~Brown, M.~Honma, T.~Mizusaki, N.~Shimizu \textit{et al.}, Phys. Rev. C  86, 051301(R) (2012). 
\bibitem{Ots19} T.~Otsuka, Y.~Tsunoda, T.~Abe , N.~Shimizu, P. ~Van~Duppen, Phys. Rev. Lett. 123, 222502 (2019)\bibitem{Woo92} J.~L.~Wood \textit{et al.}, Phys. Rep. 215, 101 (1992).
\bibitem{Hey11} K.~Heyde and J.~L.~Wood, Rev. Mod. Phys. 83, 1467 (2011).
\bibitem{Ots10} T. Otsuka, T. Suzuki, M. Honma, Y. Utsuno, N. Tsunoda, K. Tsukiyama, M. Hjorth-Jensen, Phys. Rev. Lett. 104, 012501 (2010). 
\bibitem{Ots20} T. Otsuka, A. Gade, O. Sorlin, T. Suzuki, and Y. Utsuno, Rev. Mod. Phys. 92, 015002 (2020).
\bibitem{Tsu14} Y. Tsunoda, T. Otsuka, N. Shimizu, M. Honma, Y. Utsuno, Phys. Rev. C 89, 031301(R) (2014).
\bibitem{Ots16} T. Otsuka and Y. Tsunoda, J. Phys. G: Nucl. Part. Phys. 43 024009 (2016)
\bibitem{Tog16} T. Togashi, Y. Tsunoda, T. Otsuka, and N. Shimizu, Phys. Rev. Lett. 117, 172502 (2016).
\bibitem{Mar18} B. A. Marsh \textit{et al.}, Nat. Phys. 14, 1163 (2018). 
\bibitem{Sel19} S.~Sels, T.~DayGoodacre, B.~A.~Marsh, A.~Pastore, W.~Ryssens, Y.~Tsunoda, \textit{et al.}, Phys. Rev. C 99, 044306 (2019).
\bibitem{Cri16} B. P. Crider \textit{et al.}, Phys. Lett. B763, 108 (2016).
\bibitem{Suc14} S.~Suchyta, S.~N. ~Liddick, Y.~Tsunoda, T.~Otsuka, M.~B.~Bennett,  A.~Chemey \textit{et al.}, Phys. Rev. C 89,  021301(R) (2014). 
\bibitem{Rec13} F.~Recchia, C.~J.~Chiara, R.~V.~F.~Janssens, D.~Weisshaar, A.~Gade, W.~B.~Walters \textit{et al.}, Phys. Rev. C 88, 041302(R) (2013).
\bibitem{Fla15} F. Flavigny, D.  Pauwels, D. Radulov, I. J. Darby, H. DeWitte,  J. Diriken \textit{et al.}, Phys. Rev. C 91, 034310 (2015).
\bibitem{Mue00} W. F. Mueller, B. Bruyneel, S. Franchoo, M. Huyse, J. Kurpeta, K. Kruglov \textit{et al.}, Phys. Rev. C 61,  054308 (2000).
\bibitem{Chi12} C. J. Chiara, R. Broda, W. B. Walters, R.~V.~F.~Janssens, M. Albers, M. Alcorta \textit{et al.}, Phys. Rev. C 86, 041304(R)(2012).
\bibitem{Pro15} C. J. Prokop, B. P. Crider, S. N. Liddick, A.~D.~Ayangeakaa, M.~P.~Carpenter, J.~J.~Carroll \textit{et al.}, Phys. Rev. C 92, 061302(R) (2015). 
\bibitem{Leo17} S. Leoni, B. Fornal, N. Marginean, M. Sferrazza, Y. Tsunoda, T. Otsuka \textit{et al.}, Phys. Rev. Lett. 118, 162502 (2017).
\bibitem{Hil07} S. Hilaire and M. Girod, Eur. Phys. J. A33, 237 (2007).
\bibitem{Ber91} J. F. Berger, M. Girod and D. Gogny, Comp. Phys. Comm. 63, 365 (1991).
\bibitem{Dec80} J. Decharge and D. Gogny, Phys. Rev. C 21, 1568 (1980).
\bibitem{Mol04}  P. M\"oller, A. J. Sierk, and A. Iwamoto, Phys. Rev. Lett. 92, 072501 (2004).
\bibitem{Mol09} P. M\"oller, A. J. Sierk, R. Bengtsson, H. Sagawa, and T. Ichikawa, Phys. Rev. Lett. 103, 212501 (2009).
\bibitem{Fri21} The results from the nuclear resonance fluorescence experiment confirm those discussed in the present letter, and will be published elsewhere (U. Friman-Gayer et al., to be published). 
\bibitem{Paw12} D. Pauwels, D.  Radulov, W. B. Walters, I. G. Darby,H. DeWitte, J. Diriken \textit{et al.}, Phys. Rev. C 86, 064318(2012).
\bibitem{Dar71} W. Darcey, R. Chapman, and S. Hinds, Nucl. Phys. A170, 253 (1971).
\bibitem{Bro12} R. Broda, T. Pawlat, W. Krolas, R. V. F. Janssens, S. Zhu, W. B. Walters \textit{et al.}, Phys. Rev. C 86, 064312 (2012).
\bibitem{Har92} A. Harder \textit{et al.}, Z. Phys. A 343, 7 (1992).
\bibitem{Buc16} D. Bucurescu \textit{et al.}, Nucl. Instrum. Methods Phys. Res., Sect. A 837, 1 (2016).
\bibitem{Stan20} L. Stan \textit{et al.}, to be published.
\bibitem{Mar20} N.~M\u{a}rginean \textit{et al.}, Complete low-spin spectroscopy of 64Ni by thermal n capture: tracing the evolution of shape-isomer like structures in the Ni isotopic chain, Institut Laue-Langevin (2018). https://doi.ill.fr/10.5291/ILL-DATA.3-17-36, and  N.~M\u{a}rginean \textit{et al.}, to be published.
\bibitem{Led14} C. Lederer \textit{et al.}, Phys. Rev. C 89, 025810 (2014).
\bibitem{Mic18} C. Michelagnoli \textit{et al.}, EPJ 193, 04009 (2018).
\bibitem{Str18} M. Stryjczyk, Y. Tsunoda, I. G. Darby, H. DeWitte, J. Diriken, D. V. Fedorov \textit{et al.}, Phys. Rev. C 98, 064326 (2018).
\bibitem{NUDAT} Nuclear Data Center, Brookhaven National Laboratory, https://www.nndc.bnl.gov.
\bibitem{Pas13} S. Paschalis, I. Y. Lee, A. O. Macchiavelli, C. M. Campbell, M. Cromaz, S. Gros, J. Pavan,
J. Qian, R. M. Clark, H. L. Crawford, D. Doering, P. Fallon, C. Lionberger, T. Loew, M. Petri,
T. Stezelberger, S. Zimmermann, D. C. Radford, K. Lagergren, D. Weisshaar, R. Winkler,
T. Glasmacher, J. T. Anderson, C. W. Beausang, Nucl. Instrum. Methods Phys. Res. A 709, 44 (2013).
\bibitem{Wu16} C. Y. Wu, D. Cline, A. Hayes, R. S. Flight, A. M. Melchionna, C. Zhou, I. Y. Lee, D. Swan, R. Fox, J.T. Anderson, Nucl. Instrum. Methods Phys. Res. A 814, 6 (2016).
\bibitem{Aya19} A. D. Ayangeakaa, R. V. F. Janssens, S. Zhu, D. Little, J. Henderson, C. Y. Wu, D. J. Hartley, M. Albers, K. Auranen, B. Bucher, M. P. Carpenter, P. Chowdhury, D. Cline, H. L. Crawford, P. Fallon, A. M. Forney, A. Gade, A. B. Hayes, F. G. Kondev, Krishichayan, T. Lauritsen, J. Li, A. O. Macchiavelli, D. Rhodes, D. Seweryniak, S. M. Stolze, W. B. Walters, and J. Wu, Phys. Rev. Lett. 123, 102501 (2019).
\bibitem{Czo12} T.~Czosnyka, D.~Cline, and C.~Y.~Wu, Bull. Am. Phys. Soc. 28, 745 (1983), and Gosia Manual," http://www.pas.rochester.edu/~cline/Gosia/ (2012).
\bibitem{Lit20} D.~Little, R.~V.~F.~Janssens, A.~D.~Ayangeakaa \textit{et al.}, to be published and D. Little, PhD dissertation, Univ. North Carolina at Chapel Hill, unpublished (2020).
\bibitem{Hon09} M.~Honma, T.~Otsuka, T.~Mizusaki, and M.~Hjorth-Jensen, Phys. Rev. C 80, 064323 (2009).
\bibitem{Ram01} Atomic Data and Nuclear Data Tables 78, 1-128 (2001).

\end{thebibliography}
\end{document}